\documentstyle[12pt,aasms4]{article}

\begin{document}

\title{Faint star counts in the near-infrared}

\author{J.B. Hutchings, P. B. Stetson}
\affil{Herzberg Insitute of Astrophysics,
NRC of Canada,\\ Victoria, B.C. V8X 4M6, Canada\\
john.hutchings@nrc.ca}

\author{A. Robin}
\affil{Observatiore de Besancon, BP 1615, F-25010 Besancon cedex, France\\
annie@obs-besancon.fr}

\author{T. Webb}
\affil{Dept of Astronomy, University of Toronto, Canada\\
webb@astro.utoronto.ca}

\begin{abstract}

We discuss near-infrared star counts at the Galactic pole with a view 
to guiding the
NGST and ground-based NIR cameras. Star counts from deep K-band images
from the CFHT are presented, and compared with results from the 2MASS 
survey and some Galaxy models. With appropriate corrections for detector
artifacts and galaxies, the data agree with the models down to K$\sim$18, 
but indicate a larger population of fainter red stars. There is also a
significant population of compact galaxies that extend to the observational
faint limit of K=20.5. Recent Galaxy models agree well down to K$\sim$19,
but diverge at fainter magnitudes.

\end{abstract}


\section{Introduction}

   The projected sky density of faint stars and galaxies in the near 
infrared (NIR) at high Galactic latitudes is of interest for guiding telescopes
and NIR cameras. In the case of large telescopes it is important to
choose the field of view to assure high probability of finding a guide
star, while minimising the size and cost of the guiding system. As these
instruments can guide on stars down to K$\sim$18 or fainter, it is 
necessary to know the densities at those and fainter magnitudes, since at
present we have no NIR catalogues that go this deep. 

The lowest density of stars occurs at the Galactic poles, where they are
predominantly M-type stars at these magnitudes. Thus, for guide signal 
purposes we need NIR star counts and catalogues to allow for for best 
guiding, regardless of the wavelengths of observation. The all-sky 2MASS survey
is limited to about K=16; while the HST guide star catalogue goes to fainter
V magnitudes, the NIR flux of these stars is not well known, since V-K for
M stars is in the range of 5 magnitudes, and changes steeply with spectral
type.

  In this paper we report on star counts from deep K-band images from the
CFHT IR camera, and also review other models and estimates down to K$\sim$20.

\section{Data}

  We have two deep fields from the CFHTIR camera, taken with K band filter. 
The total area covered is about 107 sq arcmin of sky. The fields are at 
Galactic coordinates l=177$^o$,  b=-50$^o$, and l=97$^o$ and b=60$^o$. 
The fields were observed as follow-up observations of objects detected 
at 850$\mu$m in two fields of the  Canada-UK Deep Submillimeter Survey 
to identify near-IR counterparts of very luminous, high-redshift, dusty
galaxies.    

The data were taken over four nights in January of 2001.  
The b=-50$^o$ field is a mosaic of four contiguous individual pointings of
integration times from 2.7 to 3.3 hours.  The b=60$^o$ field is a mosaic 
of 3 overlapping pointings of varying depth. Two pointings were observed 
for 1 hour and one pointing for 4 hours. The weather was generally worse 
in the latter and the image was therefore not as deep as the integration 
time would suggest. The final mosaic images have FWHM close to 0.9", and 
have fairly uniform depth of exposure.

The observations were processed using basic NIR techniques.  After
first removing the bias signal, sky flats were created from the program
data.  For each program image, the (dithered) images which bracketed it 
in time were combined to form a sky flat.  This sky flat was
normalized and used to correct its corresponding image.  The remaining sky
signal (a DC offset and low order structure at a few percent)  was then
removed from all the images using the IRAF task BACKGROUND, and  the
individual images  calibrated and combined to form the final image. The
HST/NICMOS faint standards were used for calibration.

Object detection and photometering was done with DAOPHOT (Stetson 1987) and
ALLSTAR (Stetson and Harris 1988).  Point-spread functions were derived from
well-exposed stars in the images, and least-squares fits of these model PSFs
were carried out to obtain relative brightnesses of the detected objects.
The SHARP and CHI indices provided by ALLSTAR were used to identify obvious
galaxies and image defects.    The conversion from
instrumental Data Numbers in the image to stellar magnitude used a
photometric zero point for the entire mosaic of images based on the
data processing and observing details.

   We obtained 2MASS K-band data for an overlapping region of sky in one 
of the CFHT fields (2MASS did not cover the other). The photometric
zero point was checked by the 2MASS magnitudes for 10 bright stars, and 
found to be consistent to within 0.1 magnitudes. The magnitude/sharpness
plots are shown in Figure 1. The exercise also 
allowed an interesting comparison of the same piece of sky
with CFHT and 2MASS (see Figure 2).

    From the two mosaic fields, we have magnitudes for some 2000 objects 
down to K$\sim$20.5 (see Figure 1). The image quality is not exceptional 
for CFHT 
but allows useful star-galaxy separation. For the cameras and telescopes
in development at present, we are interested in guiding only 
down to about K=18. At this magnitude, we have very good completeness and 
star-galaxy separation (checked by inspection). 

Completeness goes to much
fainter than this, although the asymmetrical distribution of sharpness index
among the fainter objects strongly suggests we are seeing a
population of compact galaxies that overlap the star distribution fainter
than about K=18. The limits of the star distribution are sketched
in, symmetrically about zero sharpness index value. The negative sharpness
points below the limits are noisy pixels. The boundary was established  
at faint magnitudes from the distribution of magnitude errors with
magnitude for the negative sharpness index values, because of the extra
population in the positive sharpness distribution. This was verified
by inspection of the images of objects near the boundary. The boundary was
also checked for positive sharpness objects by inspection. We also note that
the boundary values we have derived are very typical of ALLSTAR results from
many other investigations. 

   To obtain a star count free of compact galaxy contamination, the
final values used were thus from images with negative sharpness index for 
magnitudes fainter than 17.5, using the defect/star boundary 
established, and doubled to account for the real star distribution into
positive sharpness values. Table 1 shows the difference, and gives an 
indication of the compact galaxy counts, which are also of interest.

 We stress that for the stars down to magnitude K$\sim$18 - the range
of interest for guiding - we have very well established star/galaxy/blemish
separation. The investigation of fainter objects would benefit greatly
from images in different colours and with better seeing.

    The plot of star counts needs correction for Galactic latitude (and
longitude) to the pole. The factor is derived for the fields from Spagna
Fig 11 and a similar plot generated from the Bahcall-Soneira and
Besan\c{c}on models (see below). A correction value of 0.75 was used for both
CFHT fields, as an average of these models. The range of all the
individual values was about 5\%.

The total number of guide stars to our approximate guide signal limits is about
60-100, so there are some small number uncertainties. The two CFHT fields
separately give very similar star counts, and Figure 3 combines them. 
Overall, the
star counts to K=18 have error bars of 10\%. Figure 1 shows the CFHT counts
where the values for K$>$17 are free of the probable compact galaxy counts
(sharpness $>$ 0 for K$>$17.5). It still appears possible that the galaxy 
models underpredict faint stars. As the CFHT counts go fainter than 
shown in the figure, the numbers are given in the Table.

   Figure 4 shows the location in one CFHT field of stars bright enough to
guide on in the most conservative case (K=17), together with the baseline
FGS field. This was fudged to some extent: to simulate a Galactic pole
field, the magnitude limit used is brighter by the amount that Fig 1
suggests to reduce the star counts by the factor 0.75 - i.e. from 17 to
about 16.3. This probably still overestimates the guide star counts but
the figure gives a good typical example of a high latitude field and
its useable guide stars.

\section{Galaxy models}

   In Figure 3 we also  show star counts from other models. 
We discuss each in the subsections below.

\subsection{Spagna report}

   This report was written to address the NGST guide star availability,
and goes into great detail. It is available in full at the website
 http://www.ngst.nasa.gov/public/unconfigured/doc$_{-}$0422/rev$_{-}$03/NGST$_{-}$GS$_{-}$report5.pdf. In Figure 3 we show the K-band
star counts that are given for the Galactic poles in Table 9 of the
report. The model presented is based on that of Mendez and van Altena (1996)
and checked by counts from 2MASS and some smaller fields such as the
HDF. 

   The variations with Galactic latitude are shown in Figure 11 of the
report, and 
with longitude in Table 12. These values (and those from Ratnatunga 
and Bahcall 1985) were used  to correct our CFHT field counts to
the Galactic pole.

\subsection{Bahcall-Soneira model}

Ratnatunga and Bahcall (ApJ Suppl 59, 63, 1985) give star counts
in V magnitudes, in 3 B-V colour bins, for many positions in the sky,
from the Bahcall-Soneira model. The B-V bins (divided at B-V = 0.8 
and 1.3) correspond to stars of spectral type K0 and earlier, K0 - K8,
and K8 and later. Assuming mean spectral types in these bins of
F0, K4, and M3, we derive V-K colours of 4.7, 2.5, and 1.4 (Allen 2000).  
We can thus convert the star counts with V magnitude from Ratnatunga and
Bahcall to K-band magnitudes. These are shown in Figure 3. The unknown
distribution of spectral type, particularly among faint M stars, makes these
model numbers less reliable than the others shown.

These colour conversions show that all spectral types give about the same
FGS signal at the same K-band magnitude, which is consistent with the
weighted spetral response for the FGS detectors in the 0.7 to 2.5 micron
wavelength range. The J-band magnitudes are also similar for the same reason.
Thus, counting stars in the J, H, or K bands is a reasonable way to
estimate guide star availability, whereas visible magnitudes have a strong
dependence on spectral type. For this reason, catalogues in wave bands below
1 micron are not useful for NIR guide star selection.

    The plot in Fig 3 shows the sum of all stars with
K-band magnitudes for a few values, using the above corrections. 
While this model does give important detail on the
colour distribution of stars, the conversion to K-band involves some
assumptions, as noted above.

\subsection{The Besan\c{c}on model of population synthesis}

The numbers presented in Figure 3 have been computed from 
a revised version of the Besan\c{c}on model of population synthesis.
Previous versions and the main principles were described in Bienaym\'e et al. 
(1987a, 1987b) and Haywood et al. (1997). Updates of the thick disk
and spheroid parameters are given in Robin et al. (2000) and Reyl\'e \&
Robin (2001). The principles are briefly recalled below.

{\bf Model principles.}
The model is based on a semi-empirical approach, where physical constraints
and current knowledge of the formation and evolution scenario of the
Galaxy are used as a first approximation for the population synthesis. 
The model involves 4 populations (disk, thick disk, halo and bulge), 
each with specific treatment.

A standard evolution model is used to produce the populations,
based on a set of usual parameters : an initial mass function (IMF), a star
formation rate (SFR), a set of evolutionary tracks and a metallicity 
distribution. A set of IMF slopes and SFRs have been tentatively assumed 
and tested against star counts. 
The evolutionary model fixes the distribution 
of stars in the space of intrinsic
parameters : effective temperature, gravity, absolute
magnitude, mass and age. These parameters are converted into
colours in various systems through stellar atmosphere models 
corrected to fit empirical data  (Lejeune et al 1997, 1998).

The disk scale heights have been computed self-consistently from Poisson and
Boltzmann equations, using the potential
computed from the Hipparcos constraints on the local dynamical mass 
from Cr\'ez\'e et al (1998). So the disk scale heights are not free 
parameters, contrary to most galactic models, and are a function of age.

{\bf The thick disk and spheroid populations.}
In the population synthesis process,
the thick disk and spheroid populations are modeled as originating from a 
single epoch of
star formation. Bergbusch \& Vandenberg (1992) oxygen 
enhanced evolutionary tracks are used, with mean metallicities of -0.7 and 
-1.7 dex, and ages of 
11 Gyr and 14 Gyr, respectively. Their density laws and IMF have been 
determined by model-fitting to remote star counts from large data sets in the
visible and near-infrared (Robin et al. 2000, Reyl\'e et al, 2001).

\subsection{2MASS survey data}

  The 2MASS survey is not as deep as we need but does cover large
amounts of sky, and is useful as a check of the brighter star counts.
Jean-Francois Le Borgne of the CFHT group sent numbers based on the
entire 2MASS database for Galactic latitudes greater than 80$^o$. 
We obtained a smaller 2MASS dataset (800 sq arcmin) and performed the
measurements using the same DAOPHOT software as for the CFHT data
(see the plot in the lower panel of Figure 1). 

   With the same star-galaxy separation software in DAOPHOT, our counts
agree with the 2MASS point source catalogue (as also noted by Spagna).
The 2MASS-based counts from Le Borgne 
appear to include all sources in the images without star/flaw separation 
(as we verify from the raw 2MASS images) and we regard them as too high. 
The star-separated 2MASS counts are shown in Fig 3 as a single set of 
symbols, and it is clear that they do not go faint enough to be useful,
but agree with the other values as far as they go.

\section{Discussion}

    Overall, there is good agreement in Fig 3 on the star counts in
the range where the interest lies for guiding. Current performance
modelling for the NGST fine guidance camera suggests using stars down 
to the range of K=17.0 to K=17.8.  The Spagna and Robin models agree 
very closely in the range of interest in this context. They diverge at 
fainter magnitudes, but neither approaches the numbers indicated by 
the CFHT data.

    Colour information on the stars present at these
faint magnitudes is needed to determine which population is not well
reproduced by the models. Star counts being dominated by M dwarfs, it is 
probable that the discrepancy comes from the assumed IMF at low masses 
for old populations.

   The NGST Fine Guidance System has a nominal field of view of
8.4 sqare arcminutes. The star count estimates error on the worst 
number (K=17) gives a count in this FGS field between 3 and ~4. 
3 stars are the average count required for 95\% probability of finding one,
and this is the lowest number in all the above estimates. The plot allows 
estimates of the field size required once the limiting guide signal
is known, and converted to NIR band magnitude for the telescope and
detector system being used. As a guide to this, we have developed a website
that allows signal estimates for a variety of parameters. This is at
http://astrowww.phys.uvic.ca/$\sim$gwyn/ngst/phot.html.

   We will need to consider whether a new Galactic pole survey should
be undertaken to provide a catalogue of guide stars suitable for NGST.
The survey clearly needs to be done in the H or K band, since the V-K
corrections
for M stars (the most common ones at K=18) are in the range of 5 magnitudes,
and change rapidly with spectral type among such stars.

\section{Acknowledgements}

   This publication makes use of data products from the Two Micron All Sky
Survey, which is a joint project of the University of Massachusetts and
the Infrared Processing and Analysis Center/California Institute of
Technology, funded by the National Aeronautics and Space Administration 
and the National Science Foundation.

   We thank Jerry Kriss for useful discussion on some of this work, Mark 
Brodwin for participating in the CFHT observations, and Jean-Luc Beuzit
for communicating some of the earlier model results.

\clearpage
\centerline{\bf References}

Allen's astrophysical quantities, editor Arthur Cox, Springer-Varlag, 2000

Bergbusch, P.~A. \& VandenBerg, D.~A., 1992, ApJS, 81, 163

{Bienaym\'e}, O., {Robin}, A.~C., {Cr\'ez\'e}, M., 1987a, A\&A, 180, 94

{Bienaym\'e}, O., {Robin}, A.~C., {Cr\'ez\'e}, M., 1987b, A\&A, 186, 359

{Cr\'ez\'e}, M., {Chereul}, E., {Bienaym\'e}, O., {Pichon}, C.,
1998, A\&A, 329, 920

Haywood, M., Robin, A.~C., {Cr\'ez\'e}, M., 1997, A\&A, 320, 440

Lejeune T., Cusinier F., Buser R., 1997, A\&AS, 125, 229

Lejeune T., Cusinier F., Buser R., 1998, A\&AS, 130, 65

Mendez R. A., and van Altena W. F., 1996, AJ, 112, 655

Ratnatunga K. and Bahcall J.A., 1985, ApJS, 59, 63

Reyl{\' e}, C.~\& Robin, A.~C.\ 2001, A\&A, 373, 886

{Robin}, A.~C., {Reyl{\'e}}, C., {Cr{\'e}z{\'e}}, M., 2000, A\&A, 359, 103

Stetson, P. B., 1987, PASP, 99, 191

Stetson, P. B., and Harris, W. E., 1988, AJ, 96, 909

\clearpage
\centerline{\bf Captions to figures}

1. Upper: Image sharpness criterion with K band magnitude for CFHT photometry.
The dashed lines outline the locus of star/galaxy/noise separation.
To the limit of guide star magnitude (17.8), the sample is very
complete and the star/galaxy separation fully verified by inspection.
Lower: the same for the 2MASS Galactic pole images.  

2. Part of the CFHT field and the same area with 2MASS, both in K-band. 
The field size shown is close to the nominal NGST FGS field in area, although 
it is a little more elongated. The faintest stars easily visible in the 
CFHT field are K$\sim$20, and in the 2MASS field K$\sim$15. 

3. Plot of total star counts at the Galactic pole per nominal 8.4 sq 
arcmin FGS field, to K-band limiting magnitudes shown, for the various 
models and the CFHT photometry. The nominal NEA goal of 3mas is met with 
the current assumptions about the NGST telescope, and the current 
understanding of the two detectors, with the magnitude limits shown.
95\% probability of a single GS availability require an average of 3 stars
per field.

4. One of the CFHT fields outlined, with the positions of K$<$16.3 stars
(equivalent to K=17 counts at the Galactic pole star densities). The 
FGS nominal field of view is shown for comparison.

\clearpage

\begin{deluxetable}{ccc}
\tablewidth 7cm
\tablecaption{CFHTIR faint object counts per 8.4 sq arcmin}
\tablehead{\colhead{to K mag} &\colhead{\#objects} &\colhead{\#stars}}
\startdata
      18   &        7.4    &       5.1\nl
      19   &        21    &        9.4\nl
      20   &        46    &        22\nl
      20.5 &        55    &        33\nl
\enddata
\end{deluxetable}

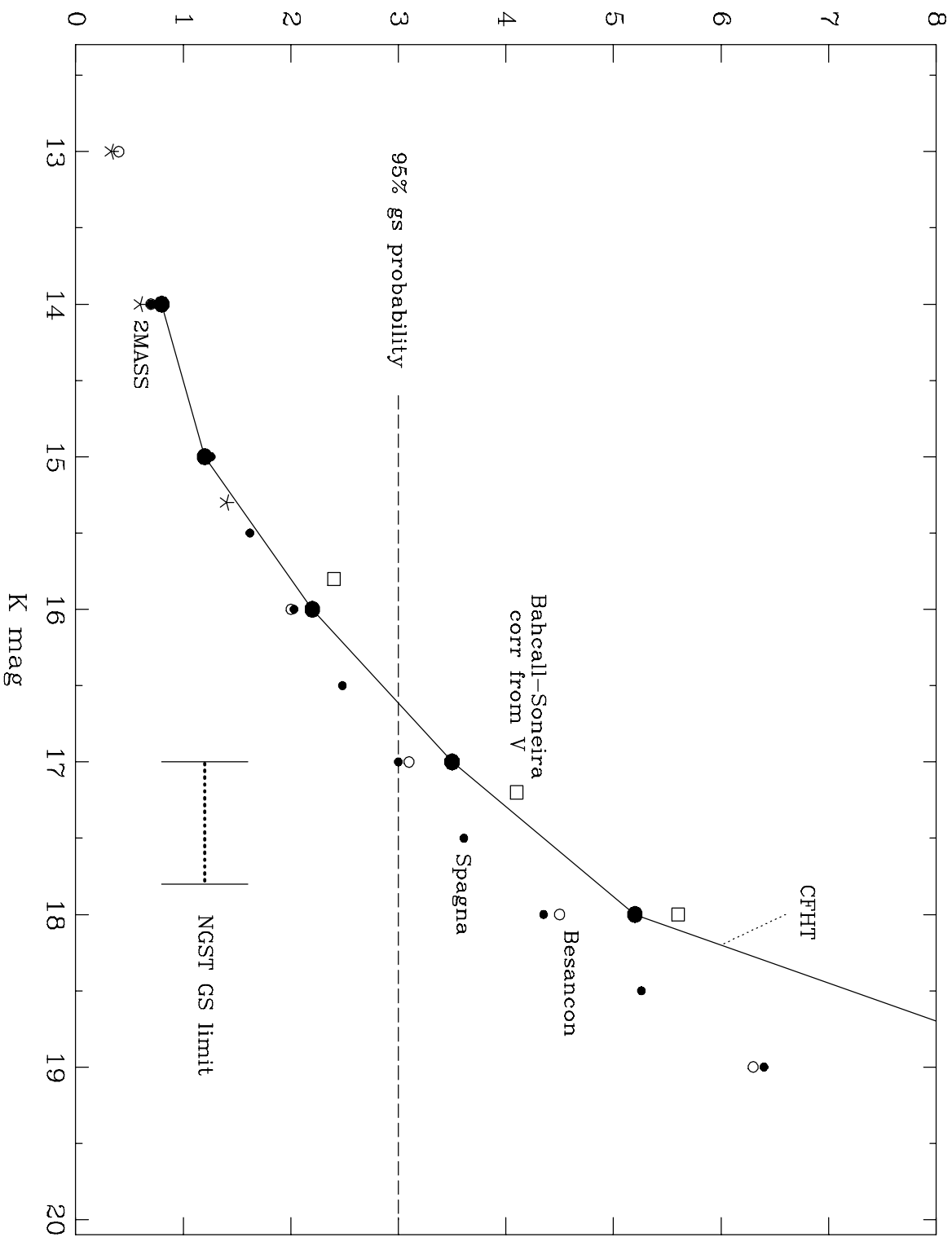
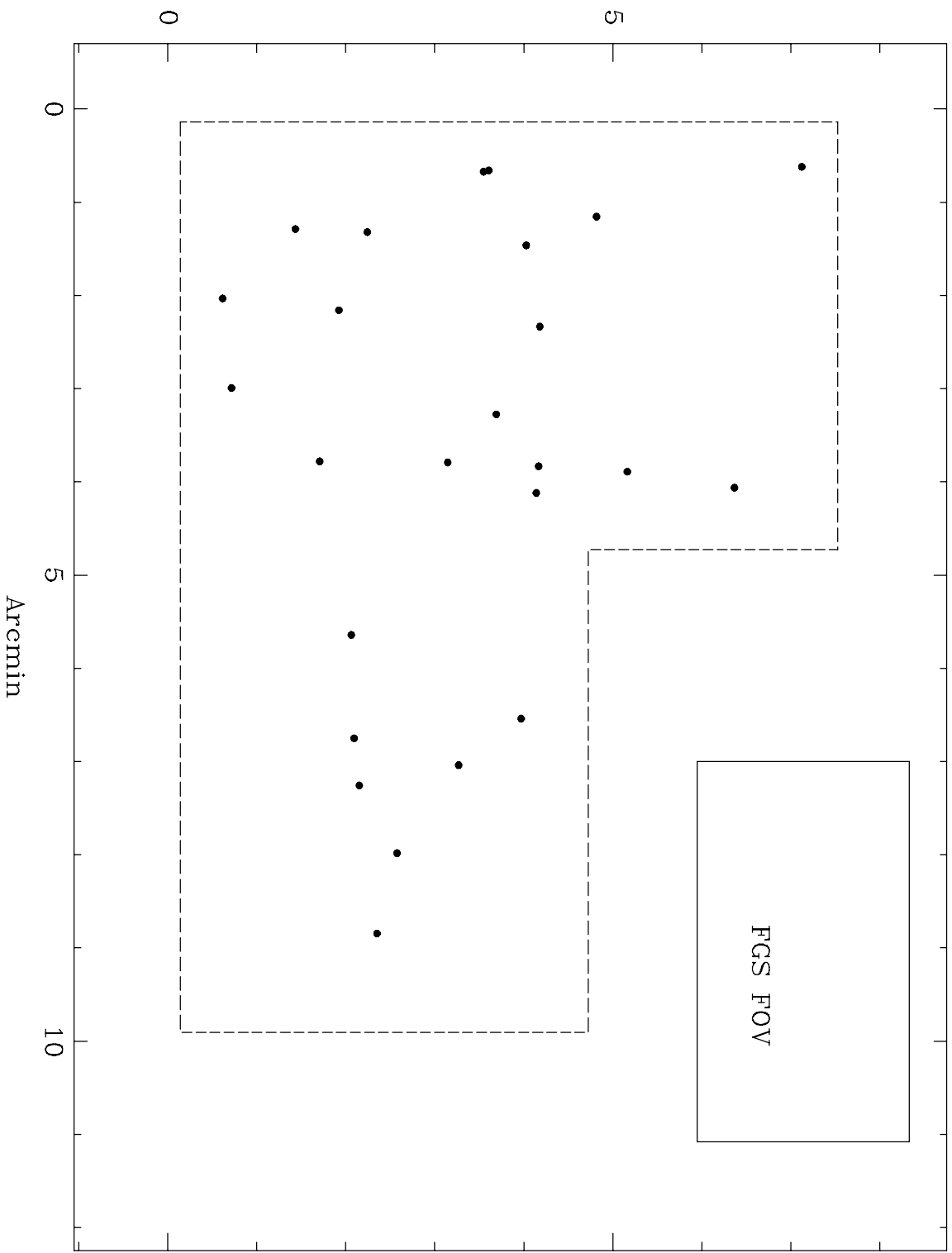
\end{document}